\documentclass[12pt]{article}
\usepackage{pdproc} 

\usepackage{graphicx}

\input{psfig}
\usepackage{epsfig}
\usepackage{pstricks}
\usepackage{axodraw}
\usepackage{amsmath, amssymb}

 \usepackage{pst-coil}
\def\f{\phi}
\def\g{\gamma}

\def\k{\kappa}

\def\m{\mu}
\def\n{\nu}
\def\o{\omega}
\def\p{\pi}

\def\s{\sigma}
\def\t{\tau}

\def\D{\Delta}

\def\G{\Gamma}

\def\L{\Lambda}

\def\ve{\varepsilon}
\def\vf{\varphi}

\def\cl{{\cal L}}

\def\bo{{\raise.15ex\hbox{\large$\Box$}}}               
\def\pr{\prod}                                          
\def\face{{\raise.2ex\hbox{$\displaystyle \bigodot$}\mskip-2.2mu \llap {$\ddot
        \smile$}}}                                      
\def\dg{\dagger}                                     
\def\beq{\begin{equation}}
\def\eeq{\end{equation}}
\def\bea{\begin{eqnarray}}
\def\eea{\end{eqnarray}}

\def\beqa{\begin{eqnarray}}
\def\eeqa{\end{eqnarray}}

\def\mt{\widetilde{m}_1}                  
\def\mb{\overline{m}}

\def\NO{\nonumber}

\def\leftrightarrowfill{$\mathsurround=0pt \mathord\leftarrow \mkern-6mu
        \cleaders\hbox{$\mkern-2mu \mathord- \mkern-2mu$}\hfill
        \mkern-6mu \mathord\rightarrow$}       
\def\dvec#1{\vbox{\ialign{##\crcr
        \leftrightarrowfill\crcr\noalign{\kern-1pt\nointerlineskip}
        $\hfil\displaystyle{#1}\hfil$\crcr}}}           

\def\pl#1#2#3{Phys.~Lett.~{\bf B {#1}} ({#2}) #3}
\def\np#1#2#3{Nucl.~Phys.~{\bf B {#1}} ({#2}) #3}
\def\prl#1#2#3{Phys.~Rev.~Lett.~{\bf #1} ({#2}) #3}
\def\pr#1#2#3{Phys.~Rev.~{\bf D {#1}} ({#2}) #3}

\def\ap#1#2#3{Ann.~of Phys.~{\bf {#1}} ({#2}) #3}

\begin{document}
{\normalsize
\begin{minipage}{5cm}
DESY 03-068
\end{minipage}}\hspace{\fill}\mbox{}\\[5ex]

\renewcommand{\thefootnote}{\alph{footnote}} 
 
\title{NEUTRINOS AND MATTER-ANTIMATTER ASYMMETRY\\ OF THE 
UNIVERSE\footnote{Presented at {\it `Neutrino Telescopes'}, Venice, March 2003}}
\author{ WILFRIED BUCHM\"ULLER}

\address{Deutsches Elektronen-Synchrotron DESY, 22603 Hamburg, Germany}

\abstract{Interactions of heavy Majorana neutrinos in the thermal phase of the
early universe may be the origin of the cosmological matter-antimatter asymmetry.
Successful baryogenesis, independent of initial conditions, is possible for
neutrino masses in the range 
$10^{-3}~\mbox{eV} \leq m_i \lesssim 0.1$~eV. Remarkably, this mass window is
consistent with the evidence for neutrino masses from oscillations.}

\normalsize\baselineskip=16.1pt

\vspace{0.8cm}
\section{Matter-Antimatter Asymmetry}

One of the main successes of the standard early-universe cosmology is the
prediction of the abundances of the light elements, D, $^3$He, $^4$He and 
$^7$Li. Agreement between theory and observation is obtained for
a certain range of the parameter $\eta_B$, the ratio of baryon density and
photon density\cite{fs02},
\begin{equation}
\eta_B^{BBN} = {n_B\over n_\g} = (2.6 - 6.2)\times 10^{-10}\;,
\end{equation}
where the present number density of photons is $n_\g \sim 400/{\rm cm}^3$. 
Since no significant amount of antimatter is observed in the universe, 
the baryon density coincides with the cosmological baryon asymmetry, 
$\eta_B =(n_B - n_{\bar{B}})/n_\g$.

The precision of measurements of the baryon asymmetry
has dramatically improved with the observation of the acoustic 
peaks in the cosmic microwave background radiation (CMB). Most recently, 
the WMAP Collaboration has measured the baryon asymmetry with a (1$\s$) standard 
error of $\sim 5\%$\cite{map03},
\begin{equation}\label{obs}
\eta_{B}^{CMB}=(6.1^{+0.3}_{-0.2})\times 10^{-10}\;.
\end{equation}

A matter-antimatter asymmetry can be dynamically generated in an expanding
universe if the particle interactions and the cosmological evolution satisfy 
Sakharov's conditions\cite{sak67}, 
\begin{itemize}
\item baryon number violation\;,
\item $C$ and $C\!P$ violation\;,
\item deviation from thermal equilibrium .
\end{itemize}
Although the baryon asymmetry is just a single number, it provides an
important relationship between the standard model of cosmology, i.e. the
expanding universe with Robertson-Walker metric, and the standard model
of particle physics as well as its extensions.

At present there exist a number of viable scenarios for baryogenesis\cite{kt90,ham02}.
They can be classified according to the different ways in which Sakharov's 
conditions are realized. In grand unified theories baryon number ($B$) and lepton
number ($L$) are broken by the interactions of gauge bosons and leptoquarks.
This is the basis of classical GUT baryogenesis\cite{kt90}.
Analogously, the lepton number violating decays of heavy Majorana neutrinos 
lead to leptogenesis\cite{fy86}. In the simplest version of
leptogenesis the initial abundance of the heavy neutrinos is generated 
by thermal processes. Alternatively,
heavy neutrinos may be produced in inflaton decays, in the reheating process
after inflation, or by brane collisions. The observed magnitude
of the baryon asymmetry can be obtained for realistic neutrino masses.

The crucial deviation from thermal equilibrium can also be realized in several
ways. One possibility is a sufficiently strong first-order electroweak phase 
transition which would make electroweak baryogenesis possible. 
For the classical GUT baryogenesis and for thermal leptogenesis the departure from
thermal equilibrium is due to the deviation of the number density of the
decaying heavy particles from the equilibrium number density.
How strong this departure from equilibrium is depends on the lifetime
of the decaying heavy particles and the cosmological evolution. 

\begin{figure}
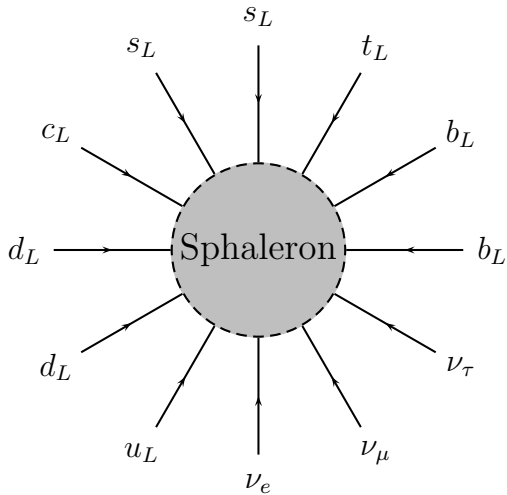

\begin{center}
\scaleboxto(7,7) {\parbox[c]{9cm}{ \begin{center}
     \pspicture*(-0.50,-2.5)(8.5,6.5)
     \psset{linecolor=lightgray}
     \qdisk(4,2){1.5cm}
     \psset{linecolor=black}
     \pscircle[linewidth=1pt,linestyle=dashed](4,2){1.5cm}
     \rput[cc]{0}(4,2){\scalebox{1.5}{Sphaleron}}
     \psline[linewidth=1pt](5.50,2.00)(7.50,2.00)
     \psline[linewidth=1pt](5.30,2.75)(7.03,3.75)
     \psline[linewidth=1pt](4.75,3.30)(5.75,5.03)
     \psline[linewidth=1pt](4.00,3.50)(4.00,5.50)
     \psline[linewidth=1pt](3.25,3.30)(2.25,5.03)
     \psline[linewidth=1pt](2.70,2.75)(0.97,3.75)
     \psline[linewidth=1pt](2.50,2.00)(0.50,2.00)
     \psline[linewidth=1pt](2.70,1.25)(0.97,0.25)
     \psline[linewidth=1pt](3.25,0.70)(2.25,-1.03)
     \psline[linewidth=1pt](4.00,0.50)(4.00,-1.50)
     \psline[linewidth=1pt](4.75,0.70)(5.75,-1.03)
     \psline[linewidth=1pt](5.30,1.25)(7.03,0.25)
     \psline[linewidth=1pt]{<-}(6.50,2.00)(6.60,2.00)
     \psline[linewidth=1pt]{<-}(6.17,3.25)(6.25,3.30)
     \psline[linewidth=1pt]{<-}(5.25,4.17)(5.30,4.25)
     \psline[linewidth=1pt]{<-}(4.00,4.50)(4.00,4.60)
     \psline[linewidth=1pt]{<-}(2.75,4.17)(2.70,4.25)
     \psline[linewidth=1pt]{<-}(1.83,3.25)(1.75,3.30)
     \psline[linewidth=1pt]{<-}(1.50,2.00)(1.40,2.00)
     \psline[linewidth=1pt]{<-}(1.83,0.75)(1.75,0.70)
     \psline[linewidth=1pt]{<-}(2.75,-0.17)(2.70,-0.25)
     \psline[linewidth=1pt]{<-}(4.00,-0.50)(4.00,-0.60)
     \psline[linewidth=1pt]{<-}(5.25,-0.17)(5.30,-0.25)
     \psline[linewidth=1pt]{<-}(6.17,0.75)(6.25,0.70)
     \rput[cc]{0}(8.00,2.00){\scalebox{1.3}{$b_L$}}
     \rput[cc]{0}(7.46,4.00){\scalebox{1.3}{$b_L$}}
     \rput[cc]{0}(6.00,5.46){\scalebox{1.3}{$t_L$}}
     \rput[cc]{0}(4.00,6.00){\scalebox{1.3}{$s_L$}}
     \rput[cc]{0}(2.00,5.46){\scalebox{1.3}{$s_L$}}
     \rput[cc]{0}(0.54,4.00){\scalebox{1.3}{$c_L$}}
     \rput[cc]{0}(0.00,2.00){\scalebox{1.3}{$d_L$}}
     \rput[cc]{0}(0.54,0.00){\scalebox{1.3}{$d_L$}}
     \rput[cc]{0}(2.00,-1.46){\scalebox{1.3}{$u_L$}}
     \rput[cc]{0}(4.00,-2.00){\scalebox{1.3}{$\nu_e$}}
     \rput[cc]{0}(6.00,-1.46){\scalebox{1.3}{$\nu_{\mu}$}}
     \rput[cc]{0}(7.46,0.00){\scalebox{1.3}{$\nu_{\tau}$}}
     \endpspicture
\end{center}}}
\end{center}
\caption{\it One of the 12-fermion processes which are in thermal 
equilibrium in the high-temperature phase of the standard model.
\label{fig_sphal}}
\end{figure}

A crucial ingredient of baryogenesis is the connection between baryon number
and lepton number in the high-temperature, symmetric phase of
the standard model. Due to the chiral nature of the weak interactions $B$ and
$L$ are not conserved\cite{tho76}. At zero temperature this has no observable 
effect due to the smallness of the weak coupling. However, as the temperature 
approaches the critical temperature $T_c$ of the electroweak phase 
transition, $B$ and $L$ violating processes come into thermal 
equilibrium\cite{krs85}. 

The rate of these processes is
related to the free energy of sphaleron-type field configurations which carry
topological charge. In the standard model they lead to an effective
interaction of all left-handed fermions\cite{tho76}  
(cf. fig.~\ref{fig_sphal}), 
\beq\label{obl}
O_{B+L} = \prod_i \left(q_{Li} q_{Li} q_{Li} l_{Li}\right)\; ,
\eeq
which violates baryon and lepton number by three units, 
\beq 
    \D B = \D L = 3\;. \label{sphal1}
\eeq
The sphaleron transition rate in the symmetric high-temperature phase
has been evaluated by combining an analytical resummation with numerical
lattice techniques\cite{bmr00}. The result is, in accord with previous 
estimates, that $B$ and $L$ violating processes are in thermal equilibrium for 
temperatures in the range
\beq 
T_{EW} \sim 100\ \mbox{GeV} < T < T_{SPH} \sim 10^{12}\ \mbox{GeV}\;.
\eeq

Sphaleron processes have a profound effect on the generation of the
cosmological baryon asymmetry.  Eq.~\ref{sphal1} suggests that any
$B+L$ asymmetry generated before the electroweak phase transition,
i.e., at temperatures $T>T_{EW}$, will be washed out. However, since
only left-handed fields couple to sphalerons, a non-zero value of
$B+L$ can persist in the high-temperature, symmetric phase if there
exists a non-vanishing $B-L$ asymmetry. An analysis of the chemical potentials
of all particle species in the high-temperature phase yields the following
relation between the baryon asymmetry and the corresponding
$L$ and $B-L$ asymmetries,
\beq\label{basic}
\langle B\rangle_T = c_S \langle B-L\rangle_T = {c_S\over c_S-1} \langle L\rangle_T\;.
\eeq
Here $c_S$ is a number ${\cal O}(1)$. In the standard model with three 
generations and one Higgs doublet one has $c_s= 28/79$. 

An important ingredient in the theory of baryogenesis is also the nature
of the electroweak transition.
A first-order phase transition yields a departure
from thermal equilibrium. Since in the standard model baryon number, $C$ and
$C\!P$ are not conserved, it is conceivable that the cosmological baryon asymmetry 
has been generated at the electroweak phase transition. Detailed studies during
the past years have shown that for Higgs masses above the present LEP bound of 
114 GeV electroweak baryogenesis is not viable, except for some 
supersymmetric extensions of the standard model\cite{qui01}.
In particular, the electroweak transition may have been just a smooth 
crossover, without any departure from thermal equilibrium. In this case it's 
sole effect in the cosmological evolution has been to switch off the $B-L$ 
changing sphaleron processes adiabatically. 

Based on the relation (\ref{basic}) between baryon and lepton number we then 
conclude that $B-L$ violation is needed to explain the cosmological baryon
asymmetry if baryogenesis took place before the electroweak transition, i.e. 
at temperatures $T > T_{EW} \sim 100$~GeV. In the standard model, as well as 
its supersymmetric version and its unified extensions based on the gauge group 
$SU(5)$, $B-L$ is a conserved quantity. Hence, no baryon asymmetry can be 
generated dynamically in these models and one has to consider extensions
with lepton number violation\footnote{In the case of Dirac neutrino masses, 
where the Yukawa couplings of right-handed neutrinos are very small, one can 
construct models where an asymmetry of lepton doublets is accompanied by an 
asymmetry of right-handed neutrinos such that the total lepton number is
conserved and $\langle B-L \rangle_T = 0$\cite{dlx00,mp02}.}.
 
\begin{figure}
\begin{center}
\scaleboxto(7,3.5){
\parbox[c]{9cm}{\input{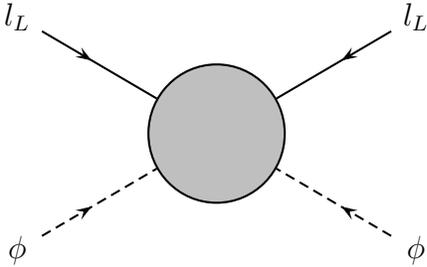}}}
\end{center}
\caption{\it Effective lepton number violating interaction.\label{fig_lept}}
\end{figure}

The remnant of lepton number violation at low energies is the appearance of an
effective $\D L=2$ interaction between lepton and Higgs fields
(cf.~fig.~\ref{fig_lept}),
\begin{equation}
\cl_{\D L=2} ={1\over 2} f_{ij}\ l^T_{Li}\vf\ C\ l_{Lj}\vf 
                  +\mbox{ h.c.}\;.\label{intl2}
\end{equation}
Such an interaction arises in particular from the 
exchange of heavy Majorana neutrinos. In the Higgs phase of the standard
model, where the Higgs field acquires a vacuum expectation value, it gives
rise to Majorana masses of the light neutrinos $\n_e$, $\n_\m$ and $\n_\t$.   

Lepton number violation appears to be necessary to understand the cosmological
baryon asymmetry. However, it can only be weak, since
otherwise any baryon asymmetry would be washed out. The interplay of these conflicting 
conditions leads to
important contraints on neutrino properties and on extensions of the standard model in 
general.   

\newpage

\section{Leptogenesis}

Lepton number is naturally violated in grand unified theories (GUTs). The 
unification of gauge couplings at high energies suggests that the standard model
gauge group is part of a larger simple group,
\begin{equation}
G_{SM} = U(1)\times SU(2)\times SU(3) \subset SU(5) \subset SO(10) \ldots\;.
\end{equation}
The simplest GUT is based on the gauge group SU(5)\cite{gg74}. Here
quarks and leptons are grouped into the multiplets,
\begin{equation}
{\bf 10} =(q_L,u_R^c,e_R^c)\;, \quad 
{\bf 5^*} = (d_R^c,l_L)\;, \quad ({\bf 1}= \n_R)\;.
\end{equation}
Unlike gauge fields, quarks and leptons are not unified in a single multiplet. 
In particular, right-handed neutrinos are not needed in SU(5) models. Since they are 
singlets with respect to SU(5), they can have Majorana 
masses $M$ which are not controlled by the Higgs mechanism.

The three SU(5) multiplets can have Yukawa interactions with two 
Higgs fields, $H_1({\bf 5})$ and $H_2({\bf 5^*})$, 
\begin{equation} 
{\cal L} = h_{uij} {\bf 10}_i {\bf 10}_j H_1({\bf 5})
          +h_{dij} {\bf 5^*}_i {\bf 10}_j H_2({\bf 5^*}) 
          +h_{\n ij} {\bf 5^*}_i {\bf 1}_j H_1({\bf 5})
          + M_{ij} {\bf 1}_i {\bf 1}_j \;.  
\end{equation}
Electroweak symmetry breaking then leads to quark and charged lepton mass matrices
and to the Dirac neutrino mass matrix $m_D = h_\n v_1$, where 
$v_1 = \langle H_1\rangle$. The Majorana mass term, which violates lepton number
($\D L = 2$), is invariant under SU(5), and the Majorana masses can therefore be much 
larger than the electroweak scale, $M \gg v$.

All quarks and leptons of one generation are unified in a single multiplet in the
GUT group SO(10)\cite{gfm75}, 
\begin{equation}
{\bf 16} = {\bf 10} + {\bf 5^*} + {\bf 1}\;.
\end{equation}
Right-handed neutrinos are now required by the fundamental gauge symmetry, and the
theory contains all ingredients needed to account for the recent evidence for 
neutrino masses and mixings. The seesaw mechanism\cite{yan79} explains the 
smallness of the light neutrino masses by the largeness of the heavy Majorana 
masses $M$. The theory predicts six Majorana neutrinos as mass eigenstates, three
heavy ($N$) and three light ($\n$),
\bea
N &\simeq& \n_R + \n_R^c \; : \qquad m_N \simeq M \; ; \\
\n &\simeq& \n_L + \n_L^c \; : \qquad m_\n = - m_D{1\over M}m_D^T \;.\label{seesaw}
\eea

In the simplest pattern of symmetry breaking, $B-L$, a subgroup of SO(10), is broken
at the unification scale $\L_{GUT}$. If Yukawa couplings of the
third generation are ${\cal O}(1)$, as it is the case for the top-quark, one finds
for the corresponding heavy and light neutrino masses: 
$M_3 \sim \L_{GUT} \sim 10^{15}$~GeV and $m_3 \sim v^2/ M_3 \sim 0.01$~eV. It is
very remarkable that the light neutrino mass $m_3$ is of the same order as
the mass differences $(\D m^2_{sol})^{1/2}$ and $(\D m^2_{atm})^{1/2}$ inferred from
neutrino oscillations. This suggests that, via the seesaw mechanism, neutrino 
masses indeed probe the grand unifiation scale! The difference of the
observed mixing patterns of quarks and leptons is a puzzle whose solution 
has to be provided by the correct GUT model. Like for quarks and charged leptons
one expects a mass hierarchy also for the right-handed neutrinos. For instance, 
if their masses scale like the up-quark masses one has 
$M_1 \sim 10^{-5} M_3 \sim 10^{10}$~GeV.

The lightest of the heavy Majorana neutrinos, $N_1$, is ideally suited to generate
the cosmological baryon asymmetry\cite{fy86}. Since it has no standard model gauge 
interactions it can naturally satisfy the out-of-equilibrium condition. $N_1$ decays 
to lepton-Higgs pairs then yield a lepton asymmetry $\langle L \rangle_T \neq 0$,
which is partially converted to a baryon asymmetry $\langle B \rangle_T \neq 0$.
The generated asymmetry is proportional to the $C\!P$ 
asymmetry\cite{fps95,crv96,bp98} in $N_1$-decays. In the case of the standard model 
with one Higgs doublet, i.e. $H_1 = H_2^* =\f$, the $C\!P$ asymmetry
is conveniently written in the following form,
\bea
\ve_1 &=& {\G(N_1 \rightarrow l \f) - \G(N_1 \rightarrow \bar{l} \bar{\f})\over 
\G(N_1 \rightarrow l \f) + \G(N_1 \rightarrow \bar{l} \bar{\f})} \NO\\
&\simeq&  {3\over 16\pi} {M_1\over (h_\n^\dg h_\n)_{11} v^2}
 \mbox{Im}\left(h_\n^\dg m_\n h_\n^*\right)_{11}\;, \label{nice}
\eea
where the seesaw mass relation (\ref{seesaw}) has been used. 

From the expression (\ref{nice})
one easily obtains a rough estimate for $\ve_1$ in terms of neutrino masses.
Assuming dominance of the largest eigenvalue of $m_\n$, phases ${\cal O}(1)$ and 
approximate cancellation of Yukawa couplings in numerator and denominator one
finds,
\begin{equation}
\ve_1 \sim {3\over 16\p}{M_1 m_3\over v^2} \sim 0.1\ {M_1\over M_3}\;,
\end{equation}
where we have again used the seesaw relation. Hence, the order of magnitude of the
$C\!P$ asymmetry is approximately given by the mass hierarchy of the heavy Majorana
neutrinos. For $M_1/M_3 \sim 10^{-5}$ ond has $\ve_1 \sim 10^{-6}$. 

Given the $C\!P$ asymmetry $\ve_1$ one obtains for the baryon asymmetry,
\bea\label{basym}
\eta_B = {n_B - n_{\bar{B}}\over n_\g} = {\k\over f} c_S \ve_1 \sim 10^{-9}\;.
\eea
Here $f \sim 10^2$ is the dilution factor which accounts for the increase of the number
of photons in a comoving volume element between baryogenesis and today. The
baryogenesis temperature is
\bea
T_B \sim M_1 \sim 10^{10}\ \mbox{GeV}\;,
\eea
and the washout factor $\k$ depends on 
the neutrino masses in a way, which will be discussed in detail in the following
chapter. Its determination requires the solution of the Boltzmann 
equations\cite{lut92,plu97}; in the estimate (\ref{basym}) we have assumed a typical 
value, $\k \sim 0.1$.
The correct value of the baryon asymmetry,  $\eta_B \sim 10^{-9}$ is then obtained
as consequence of a large hierarchy of the heavy neutrino masses, which leads to
a small $C\!P$ asymmetry, and the kinematical factors $f$ and $\k$\cite{bp96}.
The baryogenesis temperature $T_B \sim 10^{10}$~GeV corresponds to the time
$t_B \sim 10^{-26}$~s, which characterizes the next relevant epoch before recombination,
nucleosynthesis and electroweak transition.

An important question concerns the relation between leptogenesis and neutrino
mass matrices which can account for low-energy neutrino 
data\cite{alt03,buc03,fer03,kin03,moh03}.
At present we know two mass differences for the light neutrinos and we have some
information about elements of the mixing matrix $U$ in the leptonic charged current.
Since $U$ could be entirely due to mixings among the charged leptons, this does not
constrain the light neutrino mass matrix in a model independent way. The light neutrino
masses can be either quasi-degenerate or hierarchical, and they can easily be consistent
with successfull leptogenesis. In grand unified theories, due to quark-lepton 
unification, the hierarchical quark masses together with the seesaw relation require
also hierarchical heavy Majorana masses. Neglecting the mixing of charged leptons, the
constraints become very stringent and successful leptogenesis is possible only for
a very small part of parameter space\cite{afs03}. In this case the enhancement of the
$C\!P$ asymmetry for partially degenerate heavy neutrinos\cite{pil99,ery02,bra03}
plays an important role. On the other hand, in unified theories
large mixings between charged leptons easily occur. For instance, in a 
six-dimensional SO(10) model the parameters used in the above estimates
for the baryon asymmetry were recently obtained\cite{abc03}. 

\section{Quantitative Analysis and Bounds on Neutrino Masses}

Leptogenesis is a non-equilibrium process which takes place at temperatures 
$T \sim M_1$. For a decay width small compared to the Hubble parameter,
$\G_1(T) < H(T)$, heavy neutrinos are out of thermal equilibrium, otherwise they
are in thermal equilibrium. A rough estimate of the borderline between the two
regimes is given by $\G_1 = H(M_1)$\cite{kt90}. This is equivalent
to the condition that the effective neutrino mass, 
\bea\label{effm}
\mt = {(m_D^\dg m_D)_{11}\over M_1} \;,
\eea
equals the `equilibrium neutrino mass'
\begin{equation}\label{mequ}
m_* = {16\p^{5/2}\over 3\sqrt{5}} g_*^{1/2} {v^2\over M_{pl}} 
\simeq 10^{-3}~\mbox{eV}\;.
\end{equation}
Here we have used $M_{pl} = 1.2\times 10^{19}$~GeV and $g_* = 434/4$ as effective
number of degrees of freedom. For $\mt > m_*$ ( $\mt < m_*$) the heavy neutrinos of 
type $N_1$ are in (out of) thermal equilibrium at $T=M_1$. 

It is very remarkable that the equilibrium neutrino mass $m_*$ is close to the
neutrino masses suggested by neutrino oscillations,
$\sqrt{\D m^2_{\rm sol}} \simeq 8\times 10^{-3}$~eV and 
$\sqrt{\D m^2_{\rm atm}} \simeq 5\times 10^{-2}$~eV.
This suggests that it may be possible to understand the cosmological baryon 
asymmetry via leptogenesis as a process close to thermal equilibrium. Ideally,
$\D L=1$ and $\D L=2$ processes would be strong enough at temperatures above $M_1$
to keep the heavy neutrinos in thermal equilibrium and weak enough to allow
the generation of an asymmetry at temperatures below $M_1$.  

In general, the generated baryon asymmetry is the result of a competition between
production processes and washout processes which tend to erase any generated
asymmetry. Unless the heavy Majorana neutrinos are partially degenerate,
$M_{2,3}-M_1 \ll M_1$, the dominant processes are decays and inverse decays of  
$N_1$ and the usual off-shell $\D L=1$ and $\D L=2$ scatterings.  
The Boltzmann equations for leptogenesis read, 
\begin{eqnarray}\label{ke}
{dN_{N_1}\over dz} & = & -(D+S)\,(N_{N_1}-N_{N_1}^{\rm eq}) \;, \label{lg1} \\ 
{dN_{B-L}\over dz} & = & -\ve_1\,D\,(N_{N_1}-N_{N_1}^{\rm eq})-W\,N_{B-L} \;.\label{lg2}
\end{eqnarray}
Here $N_i$ are number densities, $z=M_1/T$ and $D/(Hz)$, $S/(Hz)$ and $W/(Hz)$ denote 
decay rate, $\D L=1$ scattering rate and $\D L=2$ washout rate, respectively; all rates 
are normalized to the Hubble parameter\cite{bdp02}. 

\begin{figure}[t]
\mbox{ }\hfill
\psfig{file=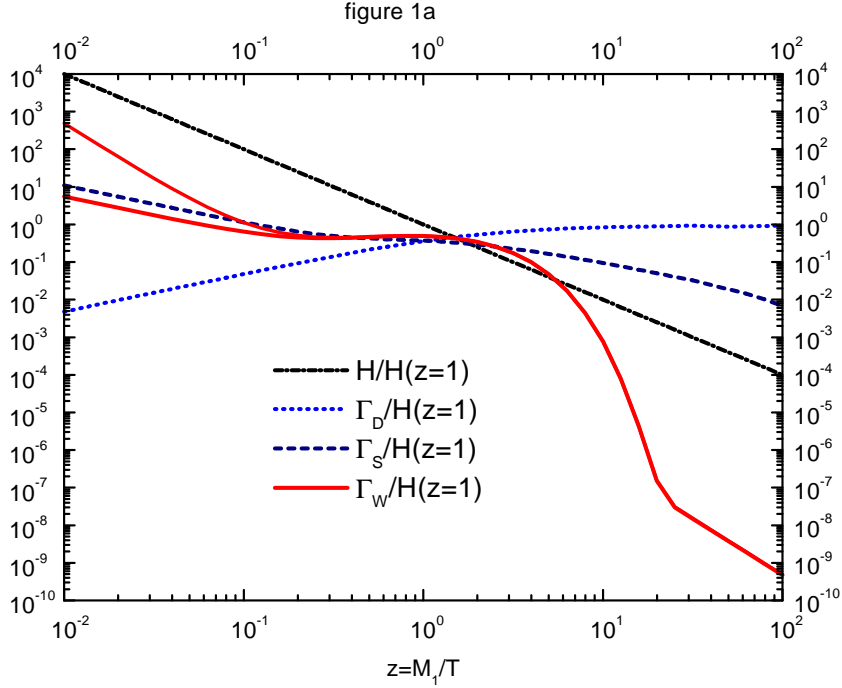,width=14cm}
\hfill\mbox{ }
\caption{\label{rates}\it Reaction rates in the thermal plasma compared with the
Hubble parameter $H$ as function of $z=T/M_1$. The neutrino parameters are 
$M_1=10^{10}\,{\rm GeV}$, $\mt=10^{-3}\,{\rm eV}$. The two branches for $\G_W$ at 
small $z$ correspond to the upper (lower) bounds $\G_W^+$ ($\G_W^-$).} 
\end{figure} 

In order to understand the dependence of the solutions on the neutrino parameters,
it is crucial to note that the rates $D/(Hz)$, $S/(Hz)$ and the resonance contribution 
$(W - \D W)/(Hz)$, are all proportional to $\mt$. One finds\cite{bdp02},
\begin{equation}\label{scaling}
D,\;  S,\;  W-\D W \; \propto \,  {M_{\rm Pl}\mt \over v^2} \;, \quad
\D W \, \propto \,  {M_{\rm Pl}M_1\, \mb^2\over v^4}  \; .
\end{equation}
Here $\mt$ is the effective neutrino mass (\ref{effm}), and $\mb$ is a quadratic mean,
\bea
\mb^2 = {\rm tr}\left(m_\n^\dg m_\n\right) = m_1^2 + m_2^2 + m_3^2 \;.
\eea
Eq.~(\ref{scaling}) implies that, as long as $\D W$ can be neglected, the generated 
lepton asymmetry is independent of $M_1$. For quasi-degenerate neutrinos, with increasing
$\mb$, the washout rate $\D W$ becomes important and eventually prevents successful
leptogenesis. This leads to the upper bound on the absolute neutrino mass scale 
discussed below.

\begin{figure}[t]
\mbox{ }\hfill
\psfig{file=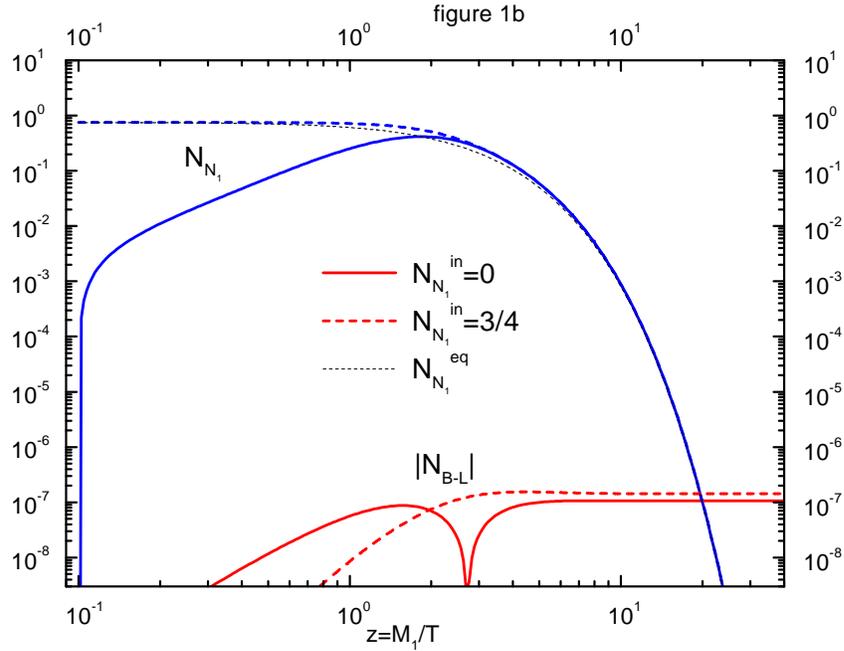,width=14cm}
\hfill\mbox{ }
\caption{\label{asymm}\it Evolution of the $N_1$ abundance and the $B-L$ asymmetry 
for $\ve_1=-10^{-6}$ and $\mb=0.05\,{\rm eV}$. The full (dashed) line corresponds 
to zero (thermal) initial $N_1$ abundance.} 
\end{figure}

The decay, scattering and washout rates are shown in fig.~\ref{rates} as functions of
$z$ for a typical set of neutrino parameters, $M_1 = 10^{10}$~GeV, $\mt = 10^{-3}$~eV,
$\mb = 0.05$~eV. All rates are of order the Hubble parameter at $z \sim 1$ where
baryogenesis takes place.
The generation of the $B-L$ asymmetry for these parameters is shown
in fig.~\ref{asymm} for $|\ve_1|=10^{-6}$ and for two different initial conditions: 
zero and thermal $N_1$ abundance. The figure demonstrates that the Yukawa
interactions are strong enough to bring the heavy neutrinos into thermal equilibrium
before leptogenesis takes place. The resulting asymmetry is in accord with observation,
$\eta_B \sim 0.01 \times N_{B-L} \sim 10^{-9}$.

Given the heavy neutrino mass $M_1$, the $C\!P$ asymmetry $\ve_1$ satisfies an
upper bound\cite{hmy02,di02}. Using this bound one can determine the maximal baryon asymmetry
$\eta^{\rm max}_{B}$ as function of the masses $\mt$, $M_1$, and $\mb$\cite{bdp02},
\bea
\eta_{B} \leq \eta^{\rm max}_{B}(\mt,M_1,\mb) \simeq 0.96\times 10^{-2} 
\ve_1^{\rm max}(\mt,M_1,\mb) \k(\mt,M_1,\mb)\;.
\eea
Requiring the maximal baryon asymmetry to be larger than the observed one, 
\bea\label{cmbcon}
\eta^{\rm max}_{B}(\mt,M_1,\mb) \geq  \eta^{CMB}_{B} \;,
\eea
yields a constraint on the neutrino mass parameters $\mt$, $M_1$ and $\mb$.

\begin{figure}
\centerline{\psfig{file=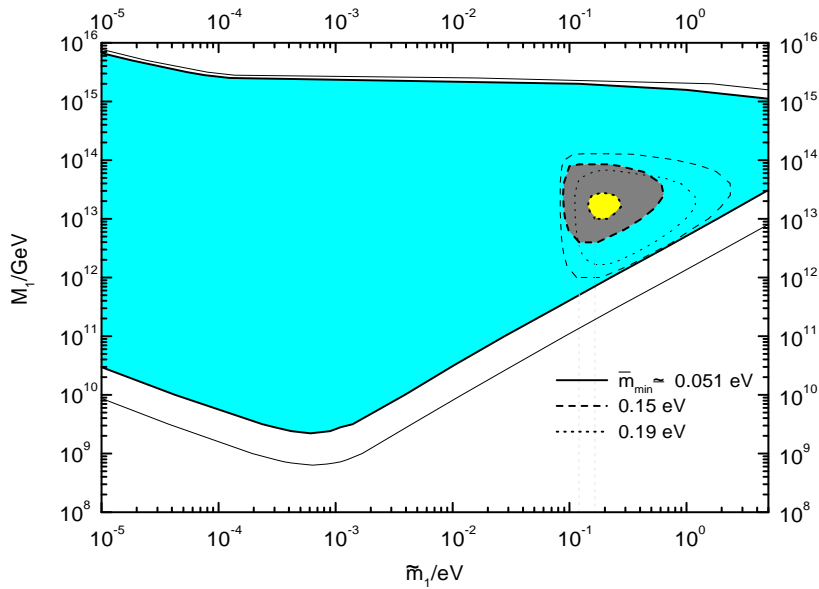,height=87mm,width=13cm}}
\caption{\label{allowed}{\small Normal hierarchy case.
Curves, in the ($\mt$-$M_1$)-plane, of constant
$\eta_{B}^{\rm max}=10^{-10}$ (thin lines) 
and $\eta_{B}^{\rm max}=3.6\times 10^{-10}$ (thick lines)
for the indicated values of $\mb$.
The filled regions for 
$\eta^{\rm max}_{B}\geq 3.6\times 10^{-10}$ are the {\it allowed regions}
from CMB. 
There is no allowed region for $\mb=0.20\,{\rm eV}$.}}
\end{figure}

The maximal $C\!P$ asymmetry as function of $\mt$, $M_1$ and $\mb$ is given 
by\cite{bdp03}
\begin{equation}\label{cpmax}
\ve_1^{\rm max} = {3\over 16\p}{M_1m_3\over v^2}
\left[1-{m_1\over m_3}\left(1+{m_3^2-m_1^2\over \mt^2}\right)^{1/2}\right]\;.
\end{equation}
For neutrino masses with normal hierarchy one has
\begin{equation} \label{nD32}
m^2_3-m^2_2  = \D m^2_{\rm atm}\; , \quad
m^2_2-m^2_1  = \D m^2_{\rm sol}\;,   
\end{equation} 
and the dependence on $\mb$ reads
\begin{eqnarray} \label{numanor1}
m_3^2 
&=& {1\over 3}\left(\mb^2 + 2\Delta m^2_{\rm atm} + \Delta m^2_{\rm sol}\right)\;, \\
\label{numanor2}
m_2^2 
&=& {1\over 3}\left(\mb^2 - \Delta m^2_{\rm atm} + \Delta m^2_{\rm sol}\right)\;, \\
\label{numanor3}
m_1^2 
&=& {1\over 3}\left(\mb^2 - \Delta m^2_{\rm atm} - 2\Delta m^2_{\rm sol}\right)\;. 
\end{eqnarray}
Note that $\ve_1 = 0$ for $\mt = m_1$. In general, one expects 
$m_1 \leq \mt \lesssim m_3$. Here the lower bound\cite{fhy02} is always true whereas
the upper bounds holds if there are no strong cancellations due to phase relations
between different elements of the neutrino mass matrix.

Using the upper bound (\ref{cpmax}) on the $C\!P$ asymmetry one can calculate the
maximal baryon asymmetry. The CMB constraint (\ref{cmbcon})
yields for each value of $\mb$ a domain in the ($\mt$-$M_1$)-plane which is allowed
by successful baryogenesis (cf.~fig.~\ref{allowed}). For $\mb < 0.20$~eV this domain
shrinks to zero. 
Using the relations (\ref{numanor1})-(\ref{numanor3}), one can easily translate this 
bound into upper limits on the individual neutrino masses,
\begin{equation}
m_1,m_2 < 0.11\,{\rm eV}\; ,\;\;\; m_3 < 0.12\,{\rm eV}\; .
\end{equation}
Note that these bounds are a factor of factor of two below the recent upper bound
of 0.23~eV obtained by WMAP\cite{map03}. For an inverted hierarchy of neutrino 
masses one finds very similar upper bounds, $m_1 < 0.11$~eV and 
$m_2, m_3 < 0.12$~eV. In a complete analysis the change of neutrino masses between
the mass scale of leptogenesis and the electroweak scale has to be included.
Generically, one expects that this will make the upper bounds on the neutrino
masses more stringent\cite{akx03}.

In a similar way one can obtain a lower bound on $M_1$, the smallest mass of the
heavy Majorana neutrinos. One finds\cite{bdp02}
\beq
M_1 > 4 \times 10^8~\mbox{GeV}\;.
\eeq
As a consequence, thermal leptogenesis requires a rather high reheating temperature,
$T_R \gtrsim T_B \sim M_1$. 

\begin{figure}
\centerline{\psfig{figure=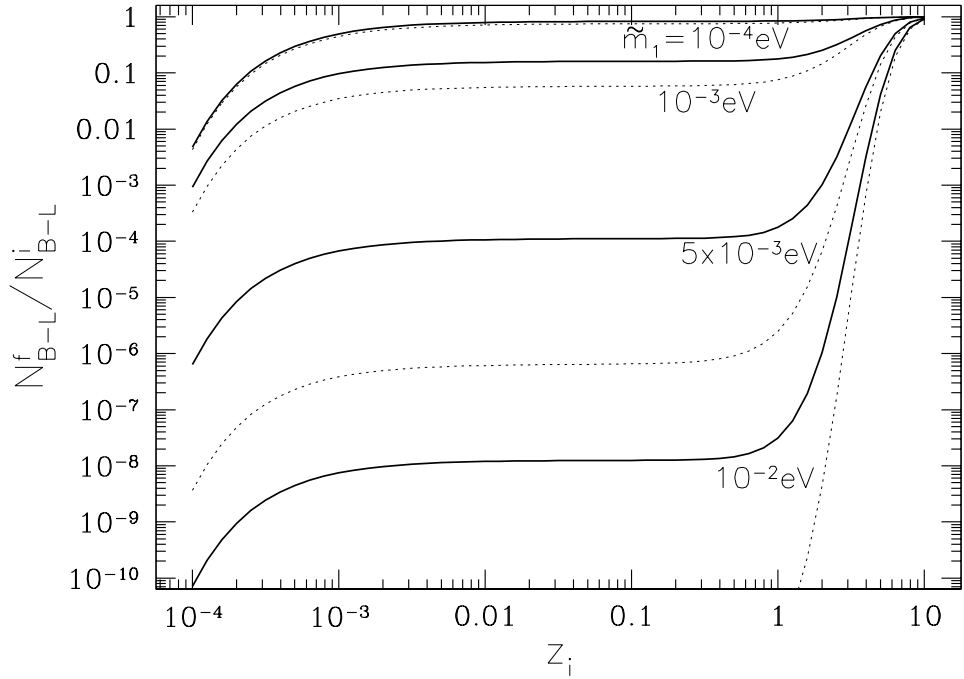,width=13cm}}
\caption{Washout factors as function of $z_i=M_1/T_{\rm i}$ without 
(full line) and with (dashed line) $N_1$-top scatterings; $M_1=10^{10}$~GeV.}
\label{wash}
\end{figure}

An important question for leptogenesis, and baryogenesis in general, is
the dependence on initial conditions. This includes the dependence on the
initial abundance of heavy Majorana neutrinos, as discussed above, and also the 
effect of an initial asymmetry which may have been generated by some other 
mechanism. It turns out that heavy Majorana neutrinos can efficiently erase
initial asymmetries.

For large initial asymmetries, one can neglect the small asymmetry generated 
through the $C\!P$ violating interactions of the heavy neutrinos, i.e. one
may set $\ve_1=0$.
The kinetic equation (\ref{lg2}) for the asymmetry then becomes
\begin{equation}\label{kew}
{dN_{B-L}\over dz} = -W\,N_{B-L} \;,
\end{equation}
where $-N_{B-L}$ is the number of lepton doublets per comoving volume.
The final $B-L$ asymmetry is then given by
\begin{equation}\label{final}
N_{B-L}^{\rm f} = \o(z_{\rm i}) N_{B-L}^{\rm i}\;,
\end{equation}
with the washout factor
\begin{equation}
\o(z_{\rm i}) = e^{-\int_{z_{\rm i}}^{\infty}\,dz\,W(z)}\;.
\end{equation} 

The result of a quantitative analysis\cite{bdp03} is shown in fig.~\ref{wash}.
The washout becomes very efficient for $\mt > m_* \simeq 10^{-3}$~eV. Already
for $\mt = 5\times 10^{-3}$~eV one has $w(z_{\rm i}) \sim 10^{-7}\ldots 10^{-5}$,
indicated by the dashed and full lines, respectively. 
Hence, an initial asymmetry several orders of magnitude larger than the presently
observed one can be reduced to a value below the one generated in leptogenesis. 
The range for $w(z_{\rm i})$
is due to a theoretical uncertainty in the treatment of $N_1$-top scatterings.
Note, that a plateau for $w(z_{\rm i})$ is reached for values of $z_{\rm i}$ 
just below one.

We conclude that leptogenesis naturally explains the observed baryon asymmetry for 
neutrino masses in the range
\bea
10^{-3}\ \mbox{eV} \leq m_i \lesssim 0.1\ \mbox{eV}\;,
\eea
almost independent of possible 
other pre-existing asymmetries. It is very remarkable that the data on solar and 
atmospheric neutrinos indicate neutrino masses precisely in this range. \\

\section{Acknowledgements}
I am grateful to Pasquale Di Bari and Michael Pl\"umacher for a fruitful collaboration
on the topic of this lecture, and I would like to thank the organizers for
arranging a lively and stimulating meeting in Venice.

\end{document}